\documentclass[aps,10pt,prl,twocolumn,superscriptaddress,showpacs,floats,floatfix,preprintnumbers,amsmath,amssymb]{revtex4-1}
\usepackage[T1]{fontenc}
\usepackage[latin9]{inputenc}
\usepackage[USenglish,american]{babel}
\usepackage{amsmath}
\usepackage{amssymb}
\usepackage{wasysym}
\usepackage{graphicx}
\usepackage{tabularx}
\usepackage{xcolor}
\usepackage{bm}          
\usepackage{pifont} 
\usepackage{dcolumn}
\usepackage{float}
\usepackage{natbib}

\newcommand{\UniRoma}{Dipartimento di Fisica, Universit\`a di Roma La Sapienza, Piazzale Aldo Moro 5, I-00185 Roma, Italy}
\newcommand{\GrazPhys}{Graz University of Technology, NAWI Graz, 8010 Graz, Austria}

\begin{document}
\title{Comment on "Pentadiamond: A Hard Carbon Allotrope of a Pentagonal Network of sp$^2$ and sp$^3$ C Atoms"}

\author{Santanu Saha}               \affiliation{\GrazPhys}
\email{santanu.saha@tugraz.at}
\author{Wolfgang von der Linden}    \affiliation{\GrazPhys}
\author{Lilia Boeri}                \affiliation{\UniRoma}
\date{\today}

\maketitle
In a recent Letter, Fujii et. al.~\cite{PentaDia} predicted a new carbon allotrope,
pentadiamond (PD), with remarkable mechanical properties: elastic moduli comparable
or larger than diamond, and negative Poisson's ratio $\mu$ = -0.241. 
The estimated Vicker's hardness($V_H$)
is 422 GPa,  $\sim$4.6 times higher than in diamond($\sim$92
GPa~\cite{avery2019predicting}),  the hardest material known to date.
PD, however was not investigated.

In the study of Avery et al.~\cite{avery2019predicting} on hardness
of different carbon allotropes  ($i$)
 $V_H$ is lower than in diamond in all cases, 
($ii$) the hardest allotropes are dominated by 
diamond and/or lonsdaleite motifs.
PD, with its record hardness and only $\sim$45.5\% $sp^3$
bonds, would be an exceptional outlier in this picture. 
What, as claimed in Ref.~[\onlinecite{PentaDia}], causes 
its extreme hardness?

Motivated by this question, we tried to reproduce the results of
Ref.~[\onlinecite{PentaDia}], recomputing the elastic properties of PD
with different approximations based on Density Functional Theory (DFT).  
Surprisingly, although we correctly reproduced the structural data,
electronic band structure and phonon dispersion of Ref.~[\onlinecite{PentaDia}],
we consistently obtained very different results for the elastic 
properties.\footnote{Pentadiamond with spacegroup Fm$\bar{3}$m has a 22
atoms primitive unit cell; with optimized lattice parameter of 9.198 \AA ~and
three inequivalent Wyckoff positions 8c (0.250, 0.250, 0.250), 32f (-0.152, -0.152,
-0.152) and 48h (0.000, -0.302, -0.302). In Ref.~[\onlinecite{PentaDia}], the 
lattice parameter was 9.195 \AA ~and the three inequivalent Wyckoff positions were 8c
(0.250, 0.250, 0.250), 32f (0.152, 0.152, 0.152) and 48h (0.198, 0.198, 0.000).}

A summary of our results is reported in 
Table \ref{tab:Moduli}. 

\begin{table}[!htb]
\centering
\caption{Calculated independent elastic constants $C_{11}$, $C_{12}$ and $C_{44}$,
Bulk modulus $B_0$, Young's modulus $E$, Shear modulus $G_0$, Vicker's hardness 
$V_H$ in GPa and dimensionless Poisson's ratio $\mu$.}
\begin{tabular}{lcccccccc}
\hline\hline 
Method & $C_{11}$ & $C_{12}$ & $C_{44}$ & $B_0$ & $E$ & $G_0$ & $V_H$ & $\mu$ \\
\cline{2-8}
  & \multicolumn{7}{c}{(GPa)} & \\
\hline 
\multicolumn{9}{l}{Pentadiamond} \\
S-S  & 509 &  94 & 142 & 237 & 404 & 166 & 23 & 0.21 \\
E-S  & 541 & 109 & 142 & 253 & 412 & 168 & 22 & 0.23 \\
Dir-PBE & 538 & 108 & 142 & 252 & 411 & 167 & 22  & 0.23\\
Dir-LDA & 568 & 121 & 143 & 270 & 424 & 171 & 21  & 0.24\\
E-S (QE) & 541 & 107 & 142 & 252 & 414 & 169 & 22 & 0.21 \\
Ref.~[\onlinecite{PentaDia}] & 1715 & -283 & 1187 & 381 & 1691 & 1113 & 422 & -0.24 \\
\hline
\multicolumn{9}{l}{Diamond} \\
S-S & 1026 & 107 & 563 & 416 & 1098 & 518 & 97 & 0.06 \\
E-S & 1052 & 133 & 565 & 439 & 1118 & 520 & 91 & 0.08 \\ 
Dir-PBE & 1049 & 121 & 567 & 430 & 1117 & 523 & 95 & 0.07 \\
Dir-LDA & 1113 & 154 & 601 & 474 & 1188 & 549 & 92 & 0.08 \\
Ref.~[\onlinecite{li2012strength}] & 1051 & 128 & 561 & 435 & 1114 & 519 & 92 & 0.07 \\
Ref.~[\onlinecite{PentaDia}] & & & & 468 & 1273 & 608 & 112 & 0.05 \\
 \hline\hline 
\end{tabular}
\label{tab:Moduli}
\end{table}

Unless otherwise specified, we employed 
the Vienna Ab-initio Simulation Package (VASP)\cite{kresse1996efficient,VASP_Kresse},
with Projected Augmented Wave(PAW) pseudopotentials \cite{PAW-VASP} for the Perdew-Burke-Ernzerhof (PBE)
exchange-correlation functional~\cite{GGA-PBE}.

The three independent cubic elastic constants
$C_{11}$, $C_{12}$ and $C_{44}$
were estimated both from the linear stress-strain (S-S) and quadratic energy-strain
relations (E-S). We created strains in the unit cell along appropriate
directions and calculated the total energy and the stress tensor after relaxation
of the internal atomic coordinates.

From the elastic constants, we obtained the bulk modulus $B_0$, Young's
modulus $E$, shear modulus $G_0$, $V_H$~\footnote{The Vicker's hardness was
based on the empirical Chen's model~\cite{chen2011modeling}, $V_{H}$ =
2$(\frac{G_0^3}{B_0^2})^{0.585}$ - 3.} and $\mu$ based
on the Voigt-Reuss-Hill approximation~\cite{hill1952elastic}. 
The same quantities were also evaluated using the built-in implementation of VASP
for the calculation of elastic constants from stress-strain relations,
based on Ref.~[\onlinecite{VASP_elastic}] (Dir).
As an independent check of our setup, the same calculations were repeated for diamond.

The elastic constants computed in the three approaches are
consistent with each other to within 5 $\%$ and, for diamond, with
literature results. On the other hand, a strong discrepancy exists between our 
results for PD and
Ref.~[\onlinecite{PentaDia}]: our elastic constants are 3-10 times
smaller, and $C_{12}$ even exhibits another sign.
As a result, our estimated $V_H$ (22 GPa) is twenty times smaller
than in Ref.~[\onlinecite{PentaDia}], and a factor four smaller than in diamond,
while the Poisson's ratio ($\mu$) is positive. We also did test runs for other hard
carbon allotropes (positive Poisson's ratio) and they match with the literature
results.

Ref~[\onlinecite{PentaDia}] did not provide sufficient computational
details to reproduce the results. 
In order to rule out  other possible
sources of discrepancy, we repeated our calculations for PD using 
VASP-PAW pseudopotentials in the Local Density Approximation~\cite{LDA-PW}(LDA) 
(Dir-LDA), and PBE norm-conserving pseudopotentials\cite{van2018pseudodojo,hamann2013optimized} 
in \textit{Quantum Espresso}, version 6.4.1\cite{QE-2017} (E-S (QE)). 
Again, the same strong discrepancy is found.

The most plausible conclusion of our tests is that the elastic constants and elastic moduli of PD reported in
Ref.~[\onlinecite{PentaDia}] are incorrect, and pentadiamond should be considered a
non-auxetic soft carbon allotrope.

{\bf Computational Details:}
For VASP calculations, we employed a kinetic energy cutoff of 800 eV, 
and a $\Gamma$-centered mesh of resolution of 2$\pi$
$\times$ 0.15 \AA$^{-1}$ for reciprocal space integration, with a Gaussian
smearing of width 0.10 eV. For QE-6.4.1, an energy cut-off of 80 Ry with Gaussian smearing of 0.02 Ry and
8 $\times$ 8 $\times$ 8 mesh on the reciprocal ($\mathbf{k}$) space was used.
This ensured a convergence of 0.4 GPa on the components of the stress tensor.
The primitive cell of both the PD and diamond was used for all the calculations.

\begin{acknowledgements}
S.S. and W.v.d.L. acknowledge computational resources from
the dCluster of the Graz University of Technology and the VSC3 of the Vienna 
University of Technology, and support through the FWF, Austrian Science Fund, Project
P 30269- N36 (Superhydra). L.B. acknowledges support from Fondo Ateneo Sapienza
2017-18 and computational Resources from CINECA, proj. Hi-TSEPH.
\end{acknowledgements}

\bibliographystyle{apsrev4-1}
\bibliography{references}
\end{document}